\documentclass[11pt,a4paper]{article}
\usepackage{jheppub}
\pdfoutput=1

\usepackage[utf8]{inputenc}
\usepackage{hyperref}
\usepackage{amsmath}
\usepackage{amsfonts}
\usepackage{amssymb}
\usepackage{graphicx}
\usepackage{color}
\usepackage{verbatim}

\newcommand{\sect}{Sec.~}

\newcommand{\app}{App.~}
\def\eq#1{Eq.~(\ref{#1})}
\def\eqno#1{(\ref{#1})}
\def\fig#1{Fig.~\ref{#1}}
\def\eqs#1{Eqs.~(\ref{#1})}

\begin{document}
\title{Improved lattice actions for supersymmetric quantum mechanics}
\author{Sebastian Schierenberg, Falk Bruckmann}
\emailAdd{sebastian.schierenberg@physik.uni-regensburg.de}
\emailAdd{falk.bruckmann@physik.uni-regensburg.de}
\affiliation{Institute for Theoretical Physics, University of Regensburg, 93040 Regensburg, Germany}

\abstract{
We analyze the Euclidean version of supersymmetric quantum mechanics on the lattice by means of a numerical path integral.  We consider two different lattice derivatives and improve the actions containing them with respect to supersymmetry by systematically adding interaction terms with non-zero extent.  To quantize this improvement, we measure boson and fermion masses and Ward identities for the naive as well as the improved models.  The masses are degenerate in all models, but the magnitude of the Ward identities decreases significantly for both derivative operators using the improved actions.  This is a clear sign that the breaking of supersymmetry due to lattice artifacts is reduced.
}

\maketitle

\section{Introduction}

Supersymmetry (SUSY) is assumed to play an important role in the search for a unified theory which incorporates all known physical interactions.  As a variety of interesting aspects of supersymmetric theories like super Yang-Mills are not accessible via perturbation theory, other approaches are required.  Promising candidates are numerical methods like the computation of path integrals on space time lattices which have been applied to quantum chromodynamics with great success.  There is, however, a fundamental problem that supersymmetric theories on discrete space-time suffer from, which is the failure of the Leibniz rule on the lattice \cite{dondi:1977, kato:2008}.  The naive ansatz for a lattice action usually is not invariant under infinitesimal translations, which it would have to be in order be fully supersymmetric.  Nevertheless, it is possible to formulate lattice theories which are not fully supersymmetric but describe the correct theory in the continuum limit.  One way to ensure this is the orbifolding procedure \cite{kaplan:2003} that allows to keep an exact subgroup of the original symmetry on the lattice.  Reviews on the progress made with this method are given in Refs.~\cite{Catterall:2009, Joseph:2011}.

In this work, we consider supersymmetric quantum mechanics (SUSYQM) on the lattice, which is a very simple model and thus a good playground to test new ideas.  For SUSYQM, degenerate boson and fermion masses are obtained at non-zero lattice spacing and with interaction by including the Wilson mass also in the bosonic derivative operator \cite{catterall:2000}.  The quadratic terms in the action are then exactly supersymmetric and the only explicit breaking comes from the interaction term which connects bosons and fermions.  This action also yields a continuum limit for the masses which is compatible with the exact value.  Further improved actions for SUSYQM are discussed in Ref.~\cite{giedt:2004}, where a lattice action preserving one exact supersymmetry, introduced in Ref.~\cite{catterall:2000}, is numerically studied.  Recently, SUSYQM on the lattice has been treated by means of a fermion loop formulation \cite{Baumgartner:2011}.  It is also possible to construct fully supersymmetric lattice theories, but this comes at the cost of a non-local action \cite{bergner:2010}, see also \cite{Bartels:1983, Kadoh:2010}.  Besides, it is not clear how to generalize this construction to lattice theories which contain non-polynomial functions of fundamental fields like gauge links.  Another approach is to keep only a remnant of the continuum SUSY on the lattice which is motivated by a blocking ansatz \cite{bergner:2009}, but no lattice action has been found for this case so far.

We will follow none of these approaches but yet another route, which is focused on finding a compromise between locality and SUSY.  To this end, we allow the interaction terms to combine fields over a finite extent (still being ultra-local) and optimize the magnitude of these combinations to minimize the breaking of translational invariance.  Due to the connection of SUSY and Poincar\'e symmetry, we assume that this also leads to an improvement of supersymmetry, which we will confirm numerically.  Contrary to most of the other approaches, we want to improve SUSY at finite lattice spacing rather than focusing on the continuum limit.  However, a combination of these improvements could be a promising ansatz for future lattice calculations.

We organize our work as follows.  In \sect\ref{sect:susyqm}, we introduce the continuum and lattice versions of supersymmetric quantum mechanics.  In \sect\ref{sect:imprsusytheory}, we  discuss our method to construct lattice models of SUSYQM with finite extent interaction terms.  Concrete lattice models are introduced in \sect\ref{sect:models}, which are classified by the two different lattice derivatives we consider.  We present our numerical results in \sect\ref{sect:numerics}, which include boson and fermion masses that are degenerate for all the models.  We furthermore show that the improved models have smaller Ward identities, which indicates a smaller breaking of SUSY.  We summarize our findings and conclude in \sect\ref{sect:summary}.

\section{Supersymmetric quantum mechanics}\label{sect:susyqm}

\subsection{Continuum}\label{sect:susyqmcont}

This section contains a very brief introduction to SUSYQM in the continuum.  The euclidean action of the theory reads
\begin{equation}\label{eq:contaction}
S = \int\mathrm d t \left[-\frac12 \chi\partial^2_t\chi + \bar\psi\partial_t\psi - \frac12 F^2 + \bar\psi\frac{\partial W}{\partial \chi}\psi - \!FW(\chi)\right]\,,
\end{equation}
with bosonic real fields $\chi$ and  $F$, and fermionic fields $\bar\psi$ and $\psi$ which depend only on the real time variable $t$.  $\partial_t$ denotes the derivative with respect to $t$ and $W(\chi)$ is a yet unspecified function called superpotential.  The action $S$  is invariant under supersymmetry transformations generated by
\begin{align}\label{eq:defgenerators}
 M = \begin{pmatrix}
      0 & 0 & 0 & 1\\
      0 & 0 & 0 & -\partial_t \\
      -\partial_t & -1 & 0 & 0 \\
      0 & 0 & 0 & 0
     \end{pmatrix}\quad\text{and}\quad
 \bar M = \begin{pmatrix}
      0 & 0 & -1 & 0\\
      0 & 0 & -\partial_t & 0 \\
      0 & 0 & 0 & 0\\
      \partial_t & -1 & 0 & 0 
      \end{pmatrix}\,,
\end{align}
which have to be understood as linear operators acting on the field multiplet $(\chi, F, \psi, \bar\psi)^T$.  The anti-commutator of these generators is proportional to the derivative $\partial_t$, i.e.                     
\begin{align}
 \{M,\bar M\} = 2\partial_t\,.
\end{align}
As the auxiliary field $F$ occurs at most quadratically in the action, it is readily integrated out, resulting in the substitution $F \to -W(\chi)$, and consequently the on-shell action
\begin{equation}
S^\text{on} = \int\mathrm d t \left[-\frac12 \chi\partial^2_t\chi +\bar\psi\partial_t\psi + \bar\psi\frac{\partial W}{\partial \chi}\psi + \frac12 W^2(\chi)\right]\,.
\end{equation}
In the on-shell formulation, the variation of the fields under SUSY transformations generated by $M$ and $\bar M$ is given by
\begin{align}
&M: \ \delta\bar\psi = 0\,, &&\delta\psi = -\epsilon\partial_t \chi + \epsilon W\,,
&&\delta\chi = \epsilon\bar\psi\,,\\
&\bar M: \ \bar\delta\bar\psi = \bar\epsilon\partial_t \chi + \bar\epsilon W\,,&&
\bar\delta\psi = 0\,,&&\bar\delta\chi = -\bar\epsilon\psi\,,
\end{align}
where $\epsilon$ and $\bar\epsilon$ are infinitesimal Grassmanian parameters.  To obtain a specific model of SUSYQM, we choose the superpotential
\begin{equation}
W(\chi) = m\chi + g\chi^3\,,
\end{equation}
with the mass $m$ and the coupling constant $g$ and obtain
\begin{align}\label{contonactionform}
S^\text{on} = \int\mathrm d t &\left[-\frac12 \chi\partial^2_t\chi + \bar\psi\partial_t\psi + m\bar\psi\psi + 3g\bar\psi\chi^2\psi+ \frac12\left(m\chi + g\chi^3\right)^2\right]\,.
\end{align}
The theory resulting from this action has a supersymmetry which is not spontaneously broken, i.e. the first bosonic and fermionic excitations have the same energies.

\subsection{Lattice}\label{sect:susyqmlatt}

A lattice version of SUSYQM is obtained by discretizing the time variable $t$.  We denote the number of lattice sites by $n$ and the lattice spacing by $a$.  All the fields acquire an index that denotes the lattice site and doubly occurring indices are implicitly summed from $0$ to $n-1$.  We consider the lattice version of the continuum action given in \eq{eq:contaction},
\begin{align}\label{eq:lattaction}
S = a\left[-\frac12 \chi_i(\nabla^2)_{ij}\chi_j + \bar\psi_i\nabla_{ij}\psi_j -\frac12 F_iF_i + \bar\psi_i\frac{\partial W_i}{\chi_j}\psi_j - F_iW_i\right]\,,
\end{align}
with the superpotential
\begin{equation}\label{eq:superpotential}
 W_i = \hat m_{ij}\chi_j + g\,T_{ijkl}\,\chi_j\chi_k\chi_l\,.
\end{equation}
The tensor $T$ and the coupling constant $g$ define the interaction term, $\nabla$ is a lattice derivative operator and $\hat m$ is a general mass term.  Each of the three will be specified later as we construct concrete lattice models. 
After integrating out the auxiliary field $F$, we obtain the lattice on-shell action
\begin{align}\label{eq:lattonaction}
S^\text{on} = a\left[-\frac12 \chi_i(\nabla)^2_{ij}\chi_j + \bar\psi_i\nabla_{ij}\psi_j + \bar\psi_i\hat m_{ij}\psi_j + 3g\,T_{ijkl}\,\bar\psi_i\chi_j\chi_k\psi_l + \frac12 W_iW_i\right]\,.
\end{align}
The discretized SUSY generators $M$ and $\bar M$ are obtained from their continuum counterparts defined in \eq{eq:defgenerators} via replacing the derivative $\partial_t$ by the lattice derivative $\nabla$.  In the on-shell formulation, they yield the variation of the fields
\begin{align}\label{eq:lattm}
&M:\ \delta\bar\psi_i = 0\,, &&\delta\psi_i = -\epsilon\nabla_{ij}\chi_j + \epsilon W_i\,,&& \delta\chi_i = \epsilon\bar\psi_i\,,\\\label{eq:lattmbar}
&\bar M:\ \bar\delta\bar\psi_i = \bar\epsilon\,\nabla_{ij}\chi_j + \bar\epsilon\,W_i\,,&&
\bar\delta\psi_i = 0\,,&&\bar\delta\chi_i = -\bar\epsilon\psi_i\,.
\end{align}
The variation of the action (\ref{eq:lattonaction}) with respect to these generators is
\begin{align}\label{eq:varaction}
&M:\ \epsilon\,\delta S = \epsilon\, g\bar\psi_i\left(3 T_{ijkl}\chi_j\chi_k\nabla_{lm}\chi_m -                     \nabla_{ij} T_{jklm}\chi_k\chi_l\chi_m\right)\,,\\
\label{eq:varactionbar}
&\bar M:\ \bar\epsilon\,\bar\delta S = \bar\epsilon\, g\psi_i\left(3 T_{ijkl}\chi_j\chi_k\nabla_{lm}\chi_m- \nabla_{ij} T_{jklm}\chi_k\chi_l\chi_m\right)\,.
\end{align}
Contrary to the continuum theory, these variations do not vanish, as $\nabla$ does not obey a Leibniz rule in general.

\section{Method to construct a SUSY-improved lattice action}\label{sect:imprsusytheory}

Due to the connection of supersymmetry and Poincar\'e symmetry, a fully supersymmetric action has to be invariant under arbitrary shifts of the fields generated by the lattice derivative $\nabla$.  We denote this property of a lattice action by \textit{continuous} translational invariance (TI).  This is an additional requirement which is not equivalent to \textit{discrete} TI, i.e. the invariance of the action under shifts of one lattice spacing.  The latter is a common property of generic lattice actions and is fulfilled by all actions considered in this work.  As a consequence of discrete TI, quadratic terms in the action also fulfill continuous TI if the lattice derivative $\nabla$ is anti-symmetric, which is usually the case.  However, for higher order interactions as characterized by the tensor $T$ in the superpotential \eqno{eq:superpotential}, continuous TI is one of the key problems of lattice SUSY. For the momentum space representations of the lattice derivative $\nabla$ and interaction tensor $T$, as defined in \app\ref{app:fourier}, this invariance is equivalent to
\begin{equation}\label{susycondform}
\big[\nabla(p) + \nabla(q) + \nabla(r) + \nabla(-p-q-r)\big]  T(p,q,r) = 0
\end{equation}
for all possible values of the momenta $p, q$ and $r$. This equation cannot be fulfilled by a local interaction term \cite{bergner:2010}.  Note that formally, \eq{susycondform} is solved by $T(p,q,r) = \delta[\nabla(p) + \nabla(q) + \nabla(r) + \nabla(-p-q-r)]$ (here, $\delta$ denotes the Kronecker symbol), as has been analyzed for $\nabla(p)\sim\sin(ap/2)$ \cite{dondi:1977, D'Adda:2010}.  However, on a finite lattice which is required for numerical simulations, such a choice of $T$ usually does not yield a sensible interaction term, because the argument of the Kronecker symbol only becomes zero for very few combinations of the momenta $p, q$ and $r$ \cite{dondi:1977}.

Our method will be to minimize the left hand side of \eq{susycondform} with a tensor $T$ that is allowed to spread over a few lattice sites.  This can be seen as a compromise between locality and supersymmetry.  The most local interaction is the point-like one called `naive',
\begin{equation}\label{mostlocalinter}
T_{ijkl}^{\tt naive} = \delta_{ijkl}\,,
\end{equation}
with $\delta_{ijkl} = \delta_{ij}\delta_{ik}\delta_{il}$.

\begin{table}
\centering
\begin{tabular}{c c c|c|c}
 $a$  & $b$ & $c$ & $L(t_{abc}^\alpha)$ & $\alpha$ \\\hline
 0 & 0 & 0 & 0 & 1\\
 0 & 0 & 1 & 3/4 & 2\\
 0 & 1 & 1 & 1 & 3\\
 1 & 1 & 2 & 2 & 4\\
 0 & 1 & 2 & 11/4 & 5\\
 0 & 0 & 2 & 3 & 6\\
 0 & 2 & 2 & 4 & 7\\
\end{tabular}
\caption{Most local contributions $t^{\alpha}_{abc}$ to the interaction tensor $T^{\tt imp}$ and their locality $L$, \eq{deflocality}, up to $L=4$.}
\label{localtab}
\end{table}

General contributions $t$ to the interaction tensor $T$ can be characterized by three lattice shifts $a$, $b$ and $c$ (due to discrete TI), and read
\begin{equation}
  (t_{abc})_{ijkl} = \text{SYMMT}\left\{\delta_{i,j+a,k+b,l+c}\right\}\,,
\end{equation}
where SYMMT$\{\}$ means a symmetrization in the indices $i$, $j$, $k$ and $l$, and under time reversal\footnote{Time reversal means a substitution of $(i, j, k, l) \to (-i, -j, -k, -l)$, where negative indices are projected back to the range $(0, n-1)$ via $-i\to -i\mod n$.}.  Due to these symmetrizations, the conditions
\begin{equation}
   0\leq a\leq b\leq c\ \text{, and}\quad a\leq c-b\,,
\end{equation}
can be imposed on the shifts without loss of generality.  As a quantity to classify such contributions, we use their degree of locality $L$, which we define by the quadratic deviation from their center of mass $X$,
\begin{equation}\label{deflocality}
 L(t_{abc}) = (0-X)^2 + (a-X)^2 + (b-X)^2 + (c-X)^2\,,\quad X = (0+a+b+c)/4\,.
\end{equation}
In Tab.~\ref{localtab} we list the most local contributions up to $L= 4$, this list can easily be extended. The ordinal number of each tensor is denoted by an upper index $\alpha$ in $t^\alpha$.  These seven tensors are exactly those, where fields are multiplied from lattice sites that are at most two sites apart, i.e. $c\leq 2$.  We restrict ourselves to these tensors because we assume that they suffice to check our method to minimize the breaking of SUSY.  However, one can in principle allow for arbitrary contributions to the interaction tensor.  In the simulations we will use an interaction tensor $T^{\tt imp}$ that is the sum over the contributions $t^\alpha$,
\begin{equation}
T^{\tt imp} = \sum_\alpha c_\alpha\ t^\alpha\,,
\end{equation}
with (real) weights $c_\alpha$ subject to the constraint
\begin{equation}\label{constraint}
\sum_\alpha c_\alpha = 1\,.
\end{equation}
The latter ensures that the superposition does not change the overall amplitude of the interaction tensor, which is given by the coupling constant $g$.

As a quantity that measures the SUSY breaking of the interaction $T$ (in combination with the lattice derivative $\nabla$), we choose 
\begin{equation}\label{eq:minform}
B (T, \nabla) = \sum_{p,q,r = -\pi/a}^{\pi/a} \Big|\big[\nabla(p) + \nabla(q) + \nabla(r) + \nabla(-p-q-r)\big]\,T(p, q, r)\Big|^2
\end{equation}
i.e. the quadratic sum over all momenta on the left hand side of \eq{susycondform}.
We shall minimize this quantity with respect to the weights $c_\alpha$.  Using a Lagrange multiplier $\lambda$ to ensure the constraint \eqno{constraint}, this can actually be done analytically, because the resulting function 
\begin{equation}
B (T, \nabla) +\lambda (\sum_\alpha c_\alpha-1) = \sum_{\alpha,\beta} Y_{\alpha\beta}c_\alpha c_\beta+\lambda\sum_{\alpha} c_\alpha-\lambda
\end{equation}
is linear and bilinear in $(c_\alpha,\lambda)$ and thus the conditions for extrema are a set of linear equations for them.  In each of the cases we consider, these have a unique solution for the $c_\alpha$.  The involved coefficients
\begin{equation}
 Y_{\alpha\beta}=\sum_{p,q,r = -\pi/a}^{\pi/a} \Big|\big[\nabla(p) + \nabla(q) + \nabla(r) + \nabla(-p-q-r)\big]\Big|^2 t^\alpha(p,q,r)t^\beta(p,q,r)
\end{equation}
are real (because the tensors $t^\alpha$ are real in momentum space), and fulfill $Y_{\alpha\beta} = Y_{\beta\alpha}$.  They depend weakly on the number of lattice sites $n$.  In the following, we therefore evaluate them in the thermodynamic $n\to\infty$ limit, where the sums over momenta are appropriately replaced by integrals.  Taking the continuum limit $a\to0$ actually makes no difference because the lattice spacing can be factorized out of $B$ for the lattice derivatives we consider, so it does not play a role in the minimization.

Because of the symmetrization, the functions $t^\alpha(p,q,r)$ consist of combinations of cosines, e.g., 
\begin{equation}
t^2(p,q,r)\propto\cos(ap) + \cos(aq) + \cos(ar) + \cos(ap+aq+ar)\,.
\end{equation}
The antisymmetric lattice derivatives are odd functions in the momenta. For the symmetric derivative to be used in \sect\ref{sect:modelswilson}, $\nabla(p)$ is a sine function and the integrals occurring in $Y_{\alpha\beta}$ can be calculated analytically.

As a rough estimation for the achieved improvement of the interaction tensor, we introduce the (non-physical) quantity
\begin{equation}\label{eq:defqual}
Q(\nabla) = \frac{B(T^{\tt imp}, \nabla)}{B(T^{\tt naive}, \nabla)}\,,
\end{equation}
where $T^{\tt imp}$ is the interaction tensor that results from a minimization of $B$.  Values of $Q$ for the improved models are provided in the respective sections.

\section{Lattice discretizations}\label{sect:models}
We consider four different discretizations based on the action given in \eq{eq:lattonaction}. These contain different kinds of lattice derivatives and interaction terms which are described in detail below.

\subsection{Naive and improved actions with the SLAC-derivative}\label{sect:modelsslac}
The derivative used in the first models is the SLAC derivative, which is defined in momentum space by
\begin{align}
\nabla_{\tt SLAC}(p) = ip\,.
\end{align}
This choice makes a Wilson mass obsolete as the SLAC operator has no doublers, which however comes at the cost of being non-local.  Therefore, the mass matrix becomes trivial,
\begin{equation}
\hat m_{ij} = m\,\delta_{ij}\,,
\end{equation}
with the bare mass $m$.

At first, we introduce the 'naive SLAC model', whose interaction tensor is given by
$T^{\tt naive}$ of \eq{mostlocalinter}.
Secondly we construct an improved model with an interaction tensor that contains contributions up to $L=4$, or equivalently, with products of fields which are up to two lattice sites apart,
\begin{equation}\label{interacttensorform}
T^{\tt imp} = \sum_{\alpha=1}^7 c_\alpha\ t^\alpha\,.
\end{equation}
The coefficients $c_\alpha$ were determined as described in \sect\ref{sect:imprsusytheory}, resulting in
\begin{align}
\nonumber
&c_1 = 0.0754\,;\ && c_2 = 0.3389\,;\ && c_3 = 0.2057\,;\ && c_4 = 0.1687\,;\\
&c_5 = 0.1597\,;\ && c_6 = 0.0421\,;\ && c_7 = 0.0095\,.\ &&
\end{align}
We refer to this choice of the interaction tensor as the 'improved SLAC model'.  This model yields an improvement of
\begin{equation}
 Q(\nabla_{\tt SLAC}) \approx 2.3\cdot 10^{-6}
\end{equation}
as defined in \eq{eq:defqual}.  

\subsection{Naive and improved actions with a Wilson mass}\label{sect:modelswilson}
In these models, we insert a symmetric difference operator into the action, which is defined by
\begin{equation}
(\nabla_{\tt symm})_{ij} = \frac1{2a}\left(\delta_{j, i+1} - \delta_{j, i-1}\right)\,.
\end{equation}
A Wilson mass has to be included to remove the doublers, so the mass matrix is
\begin{equation}
\hat m_{ij} = m\,\delta_{ij} - \frac{ra}2 \left(\delta_{i,j+1} - 2\delta_{ij} + \delta_{i,j-1}\right)\,,
\end{equation}
where we chose $r=1$ for the Wilson mass parameter.  For the 'naive Wilson model', we insert the naive interaction term as defined in \eq{mostlocalinter}.  The contributions to the improved interaction term are again the ones from \eq{interacttensorform}.  This time, the constants that are obtained by a minimization of $B(T, \nabla_{\tt symm})$ are
\begin{align}
\nonumber
&c_1 = 79/1907,\,;\ && c_2 = 448/1907\,;\ && c_3 = 318/1907\,;\ && c_4 = 432/1907\,;\\
&c_5 = 384/1907\,;\ && c_6 = 216/1907\,;\ && c_7 = 30/1907\,;\ &&
\end{align}
We call the model that results from this choice the 'improved Wilson model'. Here, we have obtained an improvement of
\begin{equation}
 Q(\nabla_{\tt symm}) \approx 5.2\cdot 10^{-4}\,.
\end{equation} 

\section{Numerical results}\label{sect:numerics}

In this section, we give the numerical results for the masses and Ward identities we have measured with the actions defined in the previous section.  For each model we have chosen a bare mass of $m=10$ and a volume of $na =1$, which ensures that the Compton wavelength of the lowest mode fits easily into the volume.  All the field configurations were obtained by a hybrid Monte Carlo algorithm similar to the one used in Ref.~\cite{bergner:2010}.

\begin{figure}
\includegraphics[width=0.48\linewidth]{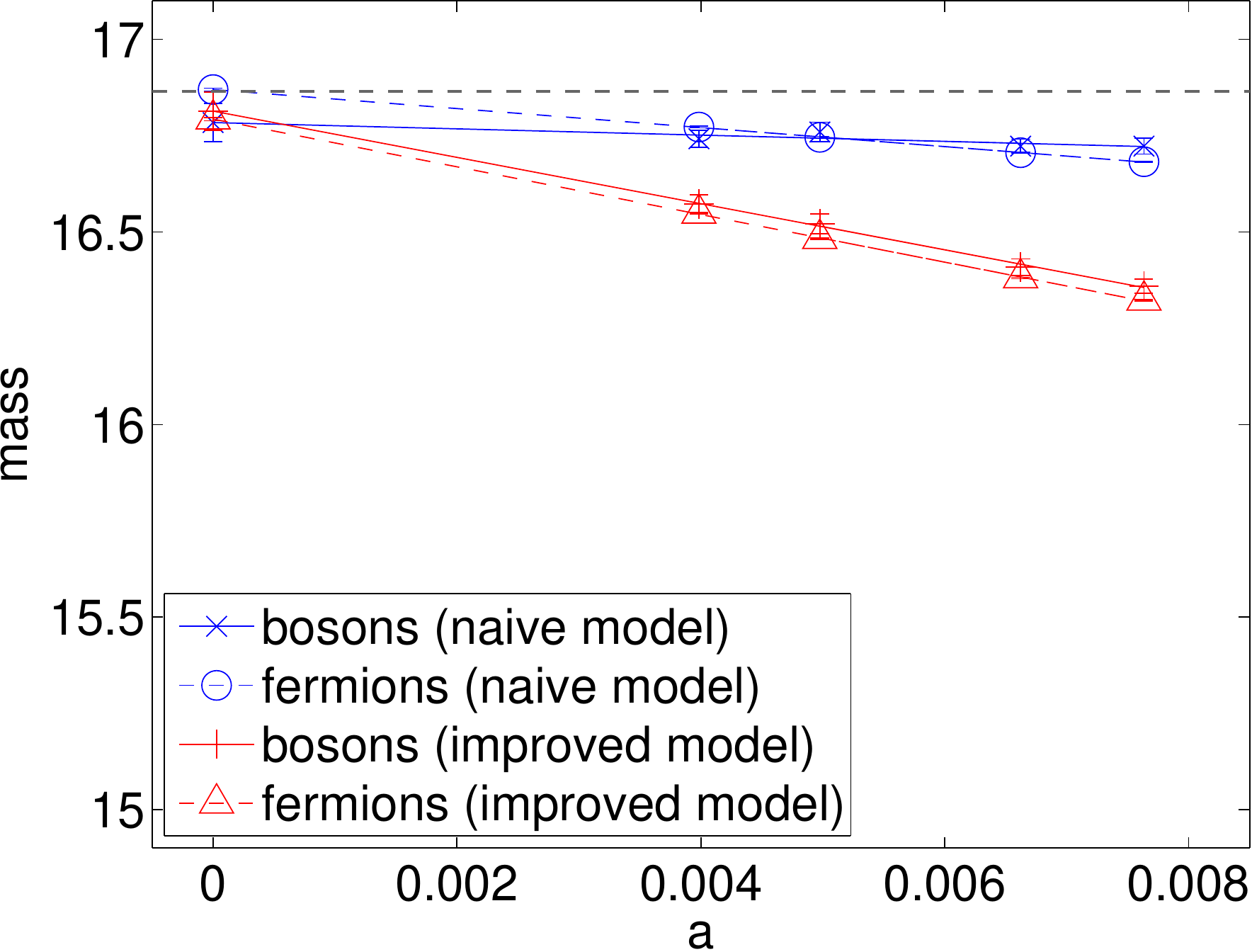}
\hspace{0.02\linewidth}
\includegraphics[width=0.48\linewidth]{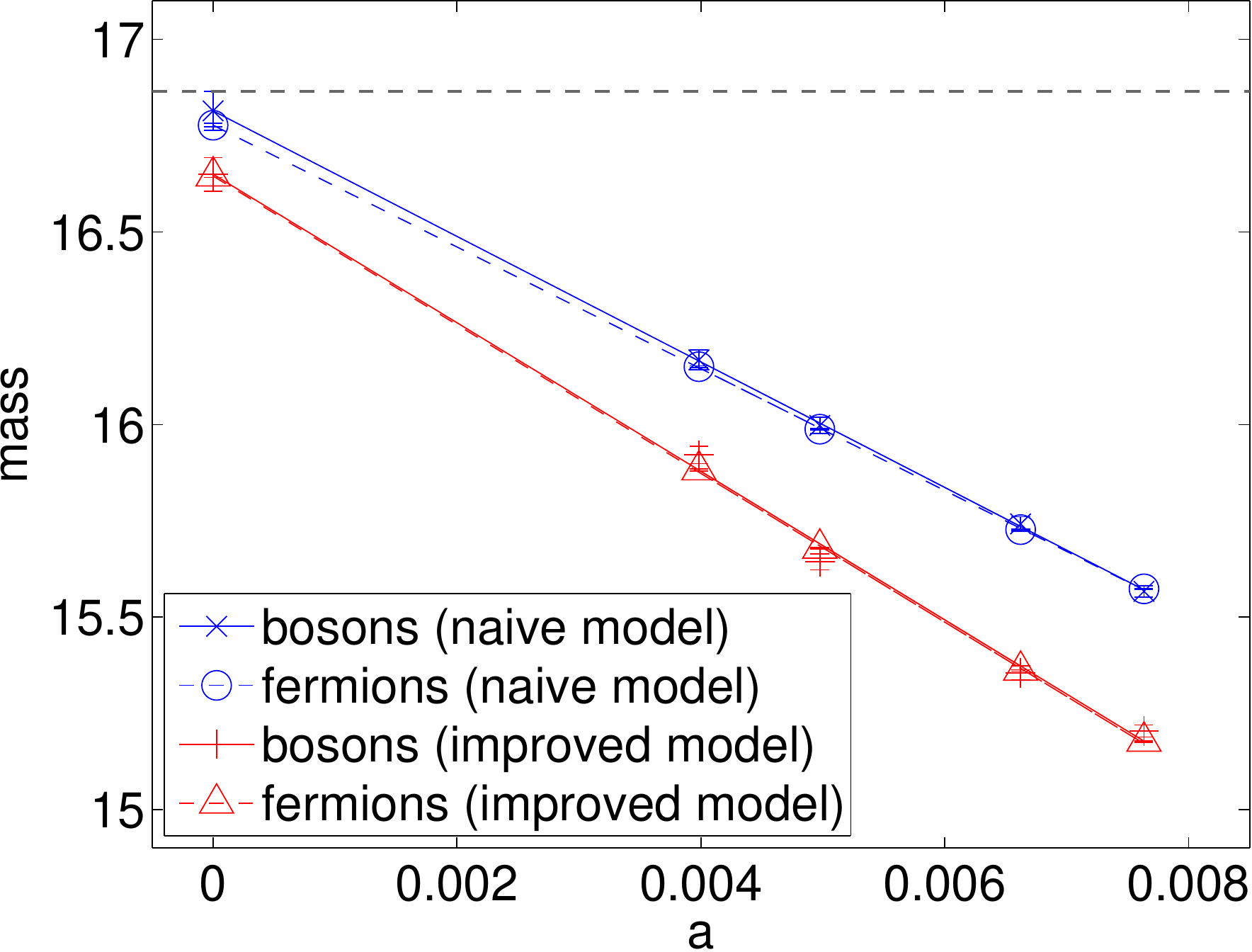}

\caption{Masses for the SLAC models (left) and Wilson models (right) for $m=10$ and $g = 100$.  The linear fits were obtained by minimized square deviation.  The horizontal dashed line is the exact continuum mass ($=16.865$).}
\label{massfigure}
\end{figure}

\subsection{Masses}

According to previous works, the naive SLAC model \cite{bergner:2008} as well as the naive Wilson model \cite{catterall:2000} defined in \sect\ref{sect:models} should have degenerate boson and fermion masses.  In order to check this and to see if it holds also for the improved models, we have calculated these masses for all the models we have defined.  This has been done by linear fits to the respective propagators in a logarithmic representation as described in Ref.~\cite{bergner:2010}.  Unlike there, we have used a simple Gaussian filter\footnote{Defined by the replacement of the propagator $\langle\bar\psi_0\psi_k\rangle\to c \sum_le^{-(k-l)^2/2}\langle\bar\psi_0\psi_l\rangle$, with $c = 1/\sum_k e^{-k^2/2}$.} to smoothen the fermionic propagator for the SLAC models, because more complicated filters have given only negligibly different results.  We have chosen a coupling constant of $g=100$, which is in the regime of strong coupling, to be able to compare our results with the previous works mentioned above, which have used the same values of $g$ and $m$.  For each lattice size and model an ensemble of $10^6$ configurations has been created, which is sufficient to obtain small errors for the propagators.  As can be seen in \fig{massfigure}, fermion and boson masses are fairly equal inside the error bars for all models and lattice spacings.  For the naive SLAC model, the masses are compatible to the ones obtained in Ref.~\cite{bergner:2008}, where the same model has been considered.  This holds also for the masses of the naive Wilson model, which have been calculated in Ref.~\cite{catterall:2000}.  The values of the respective continuum interpolations, which have been determined by a linear fit with minimized square deviation, are given in Tab.~\ref{masstab}. 

\begin{table}[t]
\centering
\begin{tabular}{c|c|c}
 \textit{model}  & $m_{\tt boson}^{\tt continuum}$ & $m_{\tt fermion}^{\tt continuum}$ \\\hline
naive SLAC & $16.784\,(94)$ & $16.870\,(5)$ \\\hline
improved SLAC & $16.813\,(50)$ & $16.793\,(4)$ \\\hline
naive Wilson & $16.815\,(50)$ & $16.778\,(5)$ \\\hline
improved Wilson & $16.650\,(43)$ & $16.645\,(4)$ \\\hline
exact & $16.865$ & $16.865$
\end{tabular}
\caption{Continuum mass extrapolations for the various actions.}
\label{masstab}
\end{table}

The naive models are closer to the exact continuum mass\footnote{The (degenerate) exact continuum mass can easily be obtained by a numerical treatment of the Hamiltonian of SUSYQM \cite{bergner:2008}.} for all values of the lattice spacing, the reason for which is presumably the larger extension of the interaction term in the improved models, as the extension is zero in the continuum.  This seems to be the price one has to pay for the improved supersymmetry.  However, the masses of the improved models also have a clear tendency towards the correct continuum mass, even if the linear fits do not hit the exact value in all cases.  We assume that combining our method with additional improvements of the action like those introduced in \cite{giedt:2004} would result in a much faster convergence of the masses to the continuum limit.

\subsection{Ward identities}

\begin{figure}[t]
\includegraphics[width=0.49\linewidth]{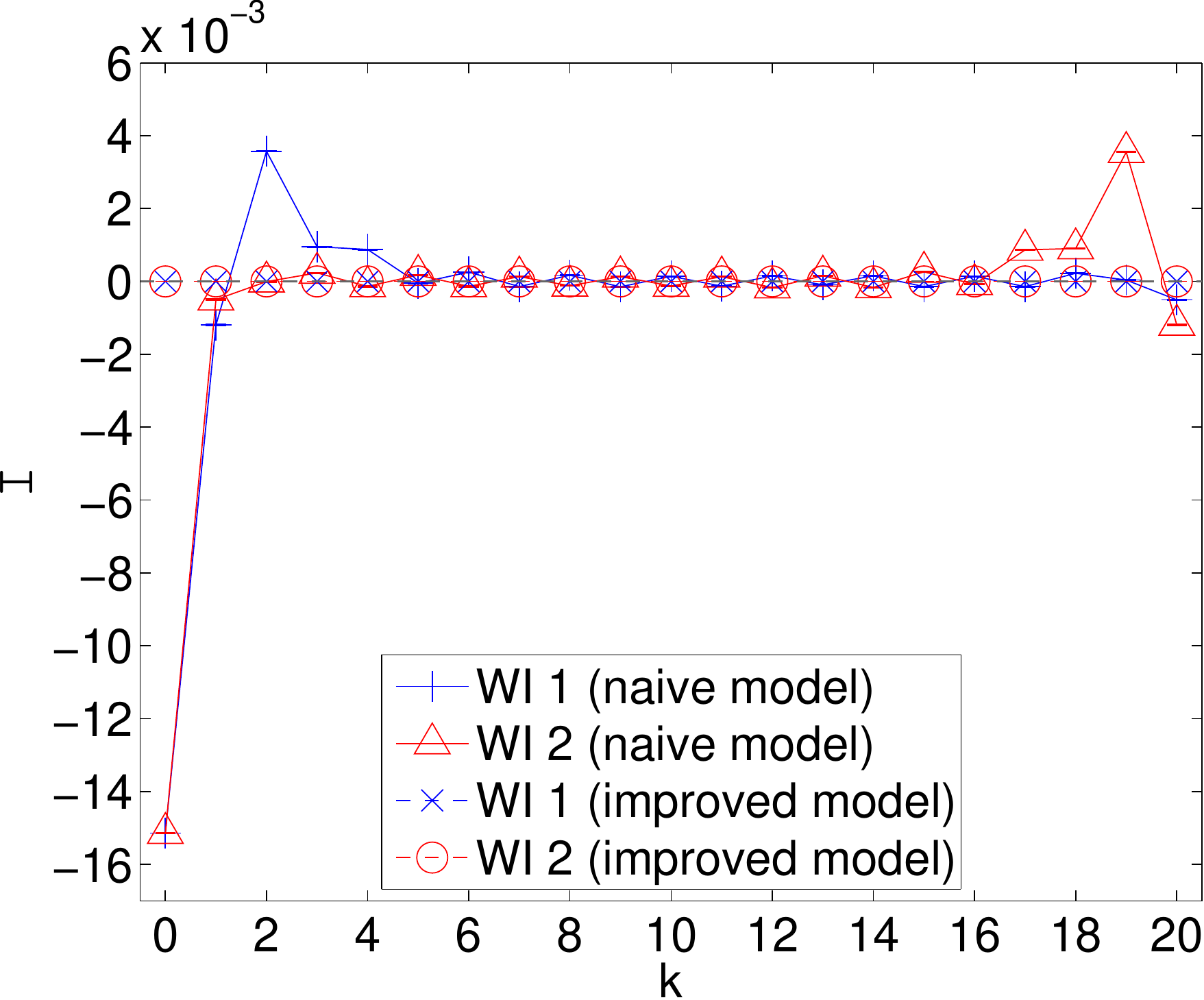}
\hspace{0.02\linewidth}
\includegraphics[width=0.49\linewidth]{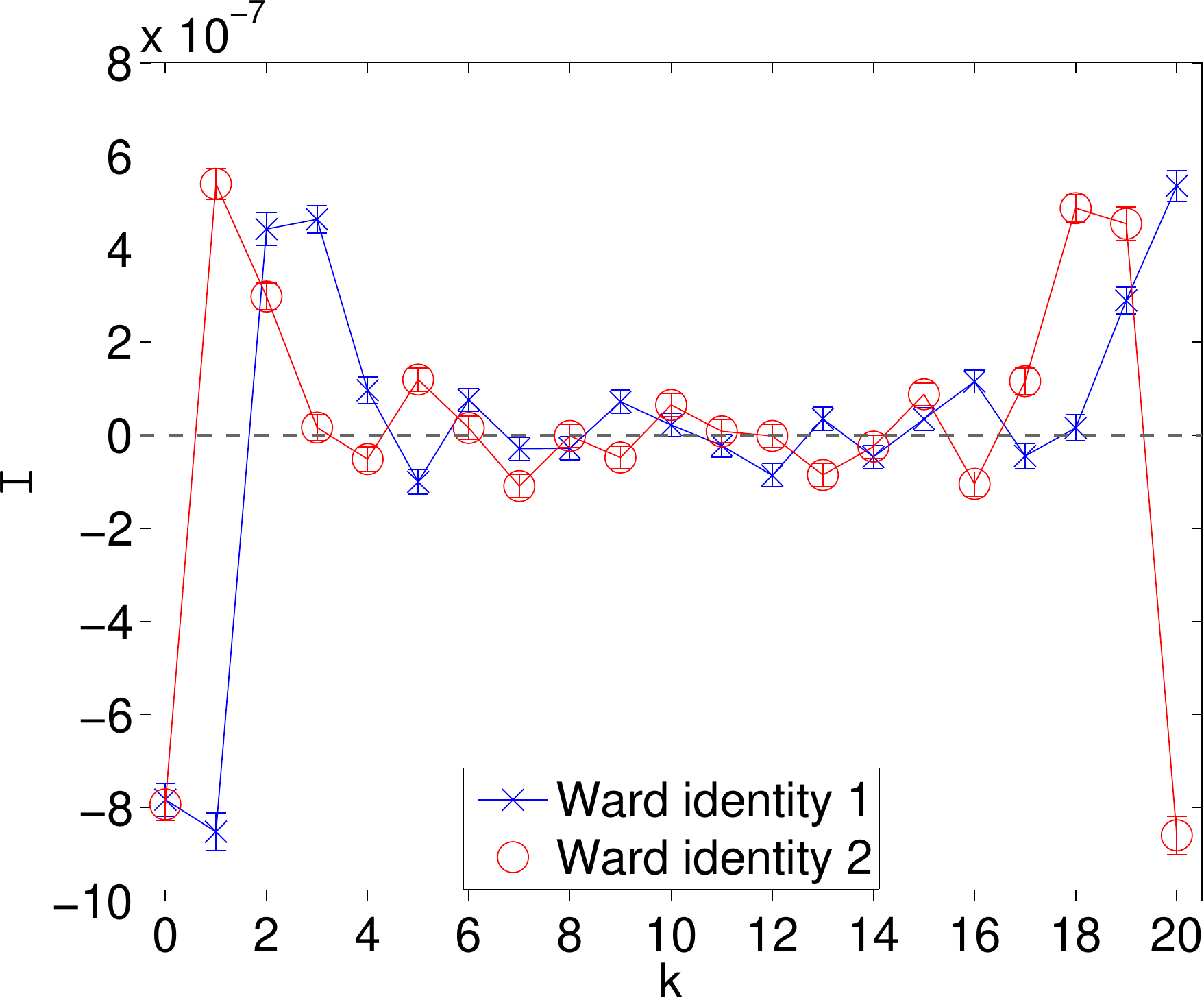}

\caption{Functions $I_{1,2}(k)$ defined in \eq{wardobs1} and \eqref{wardobs2} associated with the Ward identities for both SLAC models (left) and the improved SLAC model alone (right) for $n = 21$, $m=10$ and $g = 800$. Note the different axis scaling.}
\label{slacwardidenfigure}
\end{figure}

\begin{figure}
\includegraphics[width=0.49\linewidth]{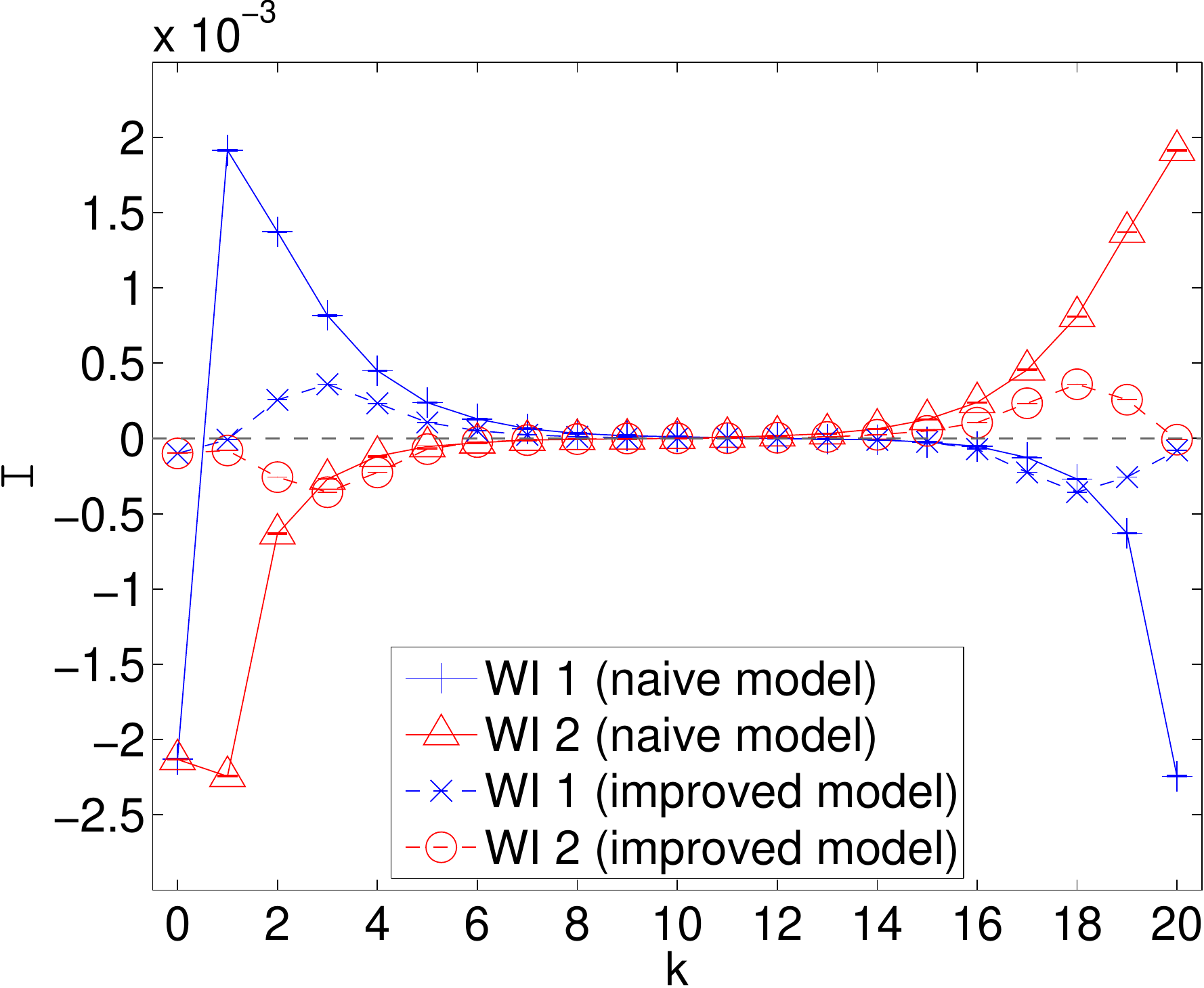}
\hspace{0.02\linewidth}
\includegraphics[width=0.47\linewidth]{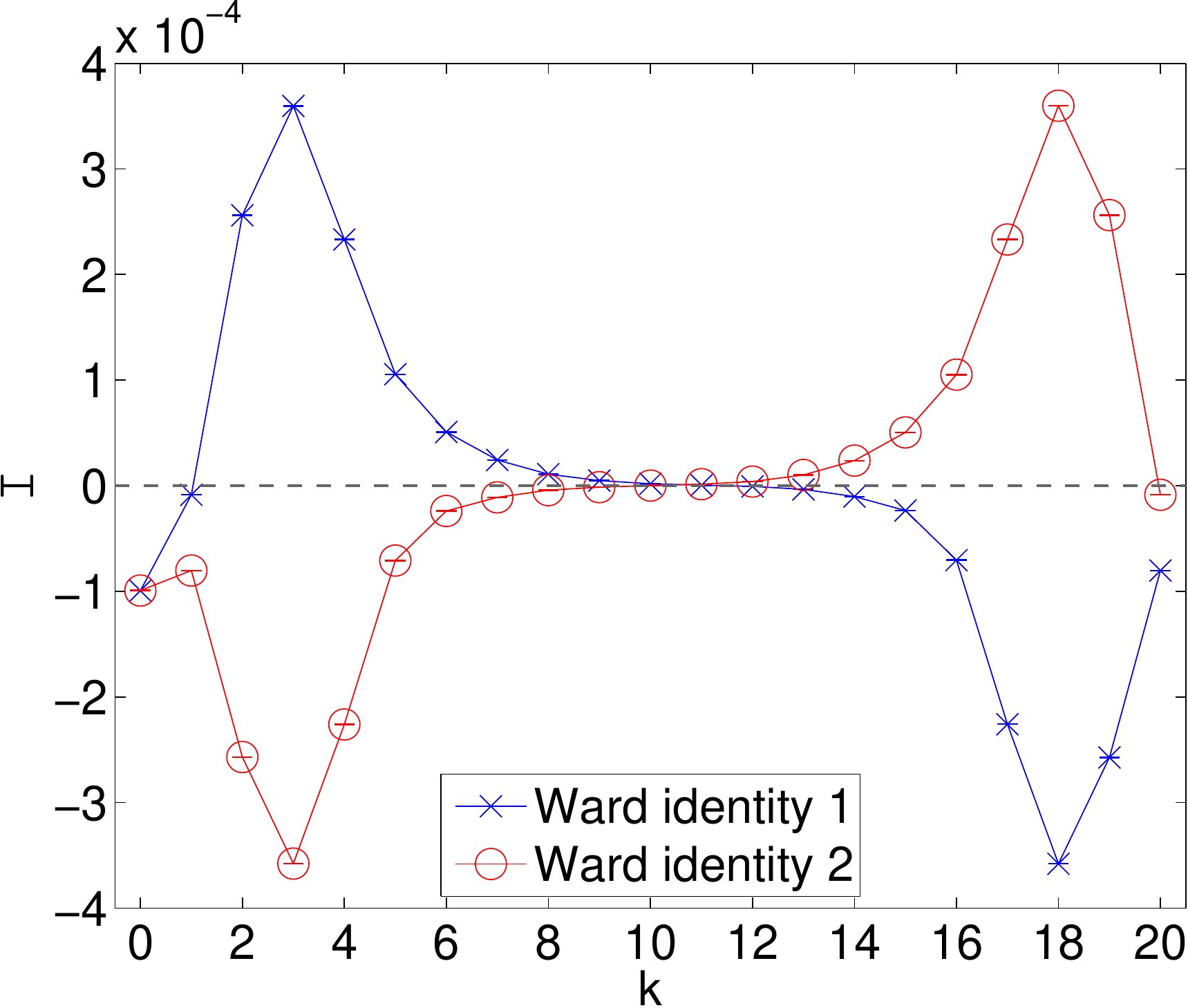}

\caption{Functions $I_{1,2}(k)$ defined in \eq{wardobs1} and \eqref{wardobs2} associated with the Ward identities for both Wilson models (left) and the improved Wilson model alone (right) for $n = 21$, $m=10$ and $g = 800$. Note the different axis scaling.}
\label{wilsonwardidenfigure}
\end{figure}

For all models, we have computed expectation values belonging to the two Ward identities (WIs) derived in \app\ref{app:wardiden},
\begin{align}\label{wardobs1}
&\mbox{WI 1:}\quad  I_1(k)=\big\langle\bar\psi_k \psi_0 + \left(-\nabla_{0j}\chi_j + W_0\right)\chi_k\big\rangle = \big\langle\delta S\,\psi_0\chi_k\big\rangle\,,\\\label{wardobs2}
&\mbox{WI 2:}\quad I_2(k)=\big\langle\bar\psi_0 \psi_k + \left(\nabla_{0j}\chi_j + W_0\right)\chi_k\big\rangle = \big\langle\bar\delta S\,\bar\psi_0\chi_k\big\rangle\,,
\end{align}
where we have set $l=0$ in \eqs{eq:wardiden1} and \eqref{eq:wardiden2} without loss of generality, because these expectation values are only functions of the difference $l-k$ due to discrete TI.  They vanish if the action is exactly supersymmetric, i.e. $\delta S = \bar\delta S = 0$, so their magnitude is a measure for the breaking of SUSY.  In Figs.~\ref{slacwardidenfigure} and \ref{wilsonwardidenfigure} we only show the expectation values related to $\delta S$ ($\bar\delta S$), i.e. the right hand sides of \eqs{wardobs1} and \eqref{wardobs2}, because these suffer from far smaller numerical errors than the other expectation values.  We have checked for each Ward identity that both expectation values coincide within the error bars.  Due to the connection between the Ward identities mentioned in \app\ref{app:wardiden}, these expectation values fulfill $I_1(k) = I_2(-k)$.  We have chosen a very strong coupling of $g=800$ and a small lattice size of $n=21$ for all the models to have a strongly broken SUSY in the naive models.

As shown in \fig{slacwardidenfigure}, the scale of the Ward identities in the improved SLAC model is \textit{smaller by four orders of magnitude} compared to the naive model.  The number of configurations is $10^7$ for the naive SLAC model and $10^8$ for the improved SLAC model.  In the latter case, better statistics have been required to get reliable results due to the small value of the Ward identities.

The same Ward identities have been computed for the Wilson models.  These are shown in \fig{wilsonwardidenfigure}, where again $10^7$ field configurations have been evaluated for the naive model and $10^8$ for the improved model.  There clearly is an improvement of about one order of magnitude, which however is much less than in the SLAC model.  One possible reason for this fact are the different values of $Q$ that were achieved, because these differ by two orders of magnitude.  Presumably, the SLAC derivative is also more suited to allow for a supersymmetric interaction term in the first place.  This assumption is backed by the fact that the only possible construction of an exactly supersymmetric (albeit non-local) lattice action with interaction requires the use of the SLAC derivative \cite{bergner:2010}.

\section{Summary}\label{sect:summary}

We have constructed lattice actions of supersymmetric quantum mechanics which are improved with respect to SUSY and have compared them to naive discretizations.  One type of models contains the SLAC derivative in the kinetic terms of the action, while the other incorporates the naive symmetric difference operator and consequently a Wilson mass term.  We have made a compromise between supersymmetry and locality by allowing the interaction term of the improved actions to connect fields within two lattice spacings or less.  Via numerical lattice simulations we have determined boson and fermion masses which have turned out to be degenerate for each model.  The Ward identities are much smaller for the improved models than for the naive ones.  We therefore conclude that the improved actions indeed have better properties with respect to supersymmetry.

The method we have developed to construct improved lattice actions can easily be generalized to include interaction terms that extend over more than two lattice sites or are polynomials of higher rank.  Furthermore, our method could also be applied to higher dimensional Wess-Zumino type models.  This would not render numerical simulations too expensive, because the interaction terms we have used are ultra-local.  Therefore, our technique could be an ingredient for the construction of supersymmetric lattice actions of more realistic models than SUSYQM, possibly in combination with other improvement methods.

\section*{Acknowledgements}

We thank Issaku Kanamori for a careful reading of the manuscript and useful remarks. This work has been supported by the DFG (BR 2872/4-2).

\bibliography{numericalsusy}
\bibliographystyle{JHEP}

\appendix

\section{Fourier transforms}\label{app:fourier}

We define the Fourier transform of a general tensor $G$ with a number of $M$ indices by
\begin{align}
 G(p_1, .., p_M) = n^{-\frac M2}\ \sum_{j_1,..,j_M=0}^{n-1} G_{j_1, .., j_M}\ e^{i a\sum_{k=1}^M p_k j_k}\,.
\end{align}
In this work, we consider only tensors which fulfill discrete TI, i.e. they are invariant under an increase of all their indices by one, in which case we obtain
\begin{align}
 G(p_1, .., p_M) = \delta_P\left(\sum_{i=1}^M p_i\right) n^{-\frac M2+1}\  \sum_{j_1,..,j_{M-1}=0}^{n-1} G_{j_1, .., j_{M-1}, 0}\ e^{i a \sum_{k=1}^{M-1} p_k j_k}\,,
\end{align}
where $\delta_P$ is a periodic $\delta$-function, defined by
\begin{align}
 \delta_P(q) = \delta\left(q\ \text{mod}\ \frac{2\pi}a\right)\,.
\end{align}
As the $\delta_P$-function constrains the sum of the momenta, we can drop one of them to arrive at the final form of the tensor $G$ in momentum space,
\begin{align}\label{eq:finalmomspactensor}
 G(p_1, .., p_{M-1}) = n^{-\frac M2+1}\ \sum_{j_1,..,j_{M-1}=0}^{n-1} G_{j_1, .., j_{M-1}, 0}\ e^{i a\sum_{k=1}^{M-1} p_k j_k}\,.
\end{align}
It is implicitly understood that the omitted momentum $p_M$ is the negative sum of all the other momenta (projected to the first Brillouin zone).

\section{Ward identities}\label{app:wardiden}

To obtain Ward identities, we consider the generating functional of a general lattice theory, defined by
\begin{equation}
Z[J] = \frac1{\mathcal N} \int D\phi\ e^{-S[\phi] + J_a \phi_a}\,,
\end{equation}
with the action $S$, the fields $\phi_a$ and the currents $J_a$. $a$ is a super-index which labels both field species as well as lattice sites. Doubly occurring super-indices are summed over all possible values they can take.  $\mathcal N$ is the partition function defined by
\begin{equation}
\mathcal N = \int D\phi\ e^{-S[\phi]}\,.
\end{equation}
An infinitesimal transformation of the fields
\begin{align}
\phi_a \to \phi_a + (\delta\phi)_a = \phi_a + G_{ab}\phi_b
\end{align}
with the generator $G$ yields
\begin{align}
\nonumber
Z[J] &\to \frac1{\mathcal N} \int D\phi \det\left[1+G\right] e^{-S[\phi] - \delta S[\phi] + J_a (\phi_a + G_{ab}\phi_b)}\\
&\approx Z[J] (1 + \mbox{tr}[G]) + \frac1{\mathcal N} \int D\phi \left(J_a G_{ab}\phi_b-\delta S[\phi]\right) e^{-S[\phi] + J_e \phi_e}\,,
\end{align}
where $\delta S$ is the variation of the action with respect to the transformation.  This has to be equal to $Z[J]$, because the transformation is equivalent to a mere change of variables in the path integral.  For a traceless $G$, we can therefore conclude
\begin{equation}
\int D\phi \left(-\delta S[\phi] + J_a G_{ab}\phi_b\right) e^{-S[\phi] + J_e \phi_e} = 0\,.
\end{equation}
Ward identities are obtained by deriving this expression with respect to currents and setting these to zero,
\begin{align}
\nonumber \frac\partial{\partial J_c} \frac\partial{\partial J_d} &\left[\int D\phi \left(-\delta S[\phi] + J_a G_{ab}\phi_b\right) e^{-S[\phi] + J_e \phi_e}\right]_{J=0}\\\nonumber
&= \int D\phi\, \big(-\delta S[\phi]\phi_c\phi_d + G_{ca}\phi_a\phi_d + \phi_c G_{da}\phi_a\big)\, e^{-S[\phi]}\\
&= \int D\phi\, \big(-\delta S[\phi]\phi_c\phi_d + (\delta\phi)_c\phi_d + \phi_c (\delta\phi)_d\big)\, e^{-S[\phi]}= 0\,.
\end{align}
In the case of SUSYQM, the generator $G$ can be either $\epsilon M$ or $\bar\epsilon\bar M$, whose actions are defined in \eqs{eq:lattm} and \eqref{eq:lattmbar}.  We specify $\phi_c = \chi_k$ and $\phi_d = \psi_l$ or $\bar\psi_l$, where $l$ and $k$ label lattice sites, to obtain the Ward identities
\begin{align}\label{eq:wardiden1}
&M:\  \big\langle\bar\psi_k \psi_l + \left(-\nabla_{lj}\chi_j + W_l\right)\chi_k \big\rangle = \big\langle\delta S\,\psi_l\chi_k\big\rangle\,,\\\label{eq:wardiden2}
&\bar M:\ \big\langle \bar\psi_l \psi_k + \left(\nabla_{lj}\chi_j + W_l\right)\chi_k\big\rangle = \big\langle\bar\delta S\,\bar\psi_l\chi_k\big\rangle\,,
\end{align}
where $\delta S$ and $\bar\delta S$ are given in \eq{eq:varaction} and \eq{eq:varactionbar}, respectively, and the average is defined by the path integral, i.e.
\begin{equation}\label{eq:pathint}
\big\langle O[\chi,\bar\psi,\psi] \big\rangle = \frac{\int D\chi D\psi D\bar\psi\ O[\chi,\bar\psi,\psi]\ e^{-S[\chi,\bar\psi,\psi]}}{\int D\chi D\psi D\bar\psi\ e^{-S[\chi,\bar\psi,\psi]}}\,.
\end{equation}
There exists a connection between the two Ward identities, which can be shown as follows.  We define
\begin{align}
 \mathcal W_{kl} = \big\langle\bar\psi_k \psi_l + \left(-\nabla_{lj}\chi_j + W_l\right)\chi_k \big\rangle,\\
\bar{\mathcal W}_{kl} = \big\langle \bar\psi_l \psi_k + \left(\nabla_{lj}\chi_j + W_l\right)\chi_k\big\rangle\,.
\end{align}
Due to the invariance of the lattice action $S$ under a time-reversal of the $\chi$-field, i.e. a replacement $\chi_i\to\chi_{-i}$ and lattice translational invariance, it is easily checked that $\mathcal W_{kl} = \bar{\mathcal W}_{lk}$.

\end{document}